# Quantifying Alignment and Quality of Graphene Nanoribbons: A Polarized Raman Spectroscopy Approach

Rimah Darawish,[1, 2] Jan Overbeck,[3, 4] Klaus Müllen,[5] Michel Calame,[3, 4, 6] Pascal Ruffieux,[1] Roman Fasel,[1, 2] and Gabriela Borin Barin[1]

[1]*Empa, Swiss Federal Laboratories for Materials Science and Technology, nanotech@surfaces Laboratory, 8600 Dübendorf, Switzerland*
[2]*Department of Chemistry, Biochemistry and Pharmaceutical Sciences, University of Bern, 3012 Bern, Switzerland*
[3]*Empa, Swiss Federal Laboratories for Materials Science and Technology, Transport at Nanoscale Interfaces Laboratory, 8600 Dübendorf, Switzerland*
[4]*Swiss Nanoscience Institute, University of Basel, 4056 Basel, Switzerland*
[5]*Max Planck Institute for Polymer Research, 55128 Mainz, Germany*
[6]*Department of Physics, University of Basel, 4056 Basel, Switzerland*

**ABSTRACT**

Graphene nanoribbons (GNRs) are atomically precise stripes of graphene with tunable electronic properties, making them promising for room-temperature switching applications like field-effect transistors (FETs). However, challenges persist in GNR processing and characterization, particularly regarding GNR alignment during device integration. In this study, we quantitatively assess the alignment and quality of 9-atom-wide armchair graphene nanoribbons (9-AGNRs) on different substrates using polarized Raman spectroscopy. Our approach incorporates an extended model that describes GNR alignment through a Gaussian distribution of angles. We not only extract the angular distribution of GNRs but also analyze polarization-independent intensity contributions to the Raman signal, providing insights into surface disorder on the growth substrate and after substrate transfer. Our findings reveal that low-coverage samples grown on Au(788) exhibit superior uniaxial alignment compared to high-coverage samples, attributed to preferential growth along step edges, as confirmed by scanning tunneling microscopy (STM). Upon substrate transfer, the alignment of low-coverage samples deteriorates, accompanied by increased surface disorder. On the other hand, high-coverage samples maintain alignment and exhibit reduced disorder on the target substrate. Our extended model enables a quantitative description of GNR alignment and quality, facilitating the development of GNR-based nanoelectronic devices.

## Introduction

Graphene nanoribbons (GNRs) are quasi-one-dimensional stripes of graphene with an intriguing set of physicochemical properties deriving from quantum confinement and related bandgap tunability[1]. The ability to tune the properties of GNRs at the atomic scale by changing their width[2–5] and edge topology [6–10] has opened up a promising avenue for their application in electronics[11–23], spintronics[24–26], and photonic devices[27–31]. The required atomic precision in GNR synthesis could only be met by a bottom-up approach based on the covalent coupling of specifically designed precursor molecules followed by cyclodehydrogenation on metallic surfaces. Since the pioneering work of Cai *et al.* in 2010[5], GNRs with various widths[2,3,30,32,33], edge topologies (armchair[34], zigzag[10], cove[35], etc. ), as well as specific edge extensions giving rise to exotic topological quantum phases, have been reported[7,6].

To explore the exciting properties of GNRs in functional devices, a substrate transfer step is necessary to transfer the GNRs from their metallic growth substrate (usually Au(111)) to semiconducting or insulating substrates suitable for digital logic applications, such as $SiO_2$/Si[11–13]. Most of the substrate transfer strategies developed so far involve aqueous solutions or the presence of polymers as a support layer, which can lead to residues or defects in the GNRs[36]. To successfully integrate GNRs into devices, GNR properties must be preserved and monitored, also upon substrate transfer, which remains one of the main bottlenecks in the development of GNR-based electronics.

Due to its speed, sensitivity, and non-destructive nature, Raman spectroscopy has emerged as one of the main techniques for probing the width, structural integrity [37–39], and even the length[38,40] of GNRs. Because it probes vibrational modes via inelastic scattering of photons, Raman spectroscopy is extremely sensitive to geometric structure within molecules[41]. This makes it a powerful technique to characterize GNRs from their growth conditions under ultrahigh vacuum (UHV) to their device integration[11–15]. Owing to the largely anisotropic properties of GNRs, polarization-dependent Raman spectroscopy is key to characterizing the overall alignment of such quasi-1D structures[29,38]. From a device perspective, the degree of GNR alignment is an extremely important feature[12,13,15,16]. For FETs, for example, the device yield is significantly improved when transferring GNRs in the same orientation as the pre-patterned source and drain contacts[13,40]. Similar to optoelectronic devices, where absorption and emission of light are most efficient for polarization along the GNR axis[30,27].

To characterize the degree of GNR alignment, previous studies used the Raman polarization anisotropy ($P$) which is defined as $P = (I_\parallel - I_\perp)/ (I_\parallel + I_\perp)$, where $I_\parallel$ and $I_\perp$ are the Raman intensities measured with polarization along and perpendicular to the GNR axis, respectively. Polarization anisotropy $P =1$ thus corresponds to perfect uniaxial alignment of GNRs, whereas $P = 0$ indicates random GNR orientation with no preferential direction of alignment[29,38,27]. These studies demonstrated preservation of the GNRs' overall degree of alignment upon substrate transfer for the case of a complete monolayer of 7- and 9-atom-wide armchair GNRs (7-AGNR and 9-AGNR, respectively), with $P > 0.7$-$0.8$[29,38,27]. Despite the fact that GNR alignment within a full monolayer was successfully preserved, GNR transport properties were heavily deteriorated by GNR-GNR bundling, which resulted in hopping transport (inter-GNR conduction) perpendicular to the source and drain contacts in FETs with channel lengths of 60 nm[15]. Theoretical studies on semiconducting carbon nanotubes (SCNTs) also showed that, as the separation between CNTs decreases, the CNT-FET characteristics are degraded. This degradation was associated with charge screening between neighboring CNTs in the channel and to Schottky barriers at the CNT/metal contact interface [42]. It is thus clear that the alignment and distribution of GNRs are important characteristics that impact device performance. The Raman polarization anisotropy ($P$) approach used in previous studies provides the overall alignment by only taking into account two data points: the intensity parallel and perpendicular to the assumed GNR alignment direction, which, however, is generally not precisely known. Therefore, this method is limited in its ability to provide a complete characterization of the angular distribution of the GNRs' long axis relative to an arbitrary in-plane axis, and it does not consider any other contributions that may arise from polarization-independent Raman intensities.

In this work, we investigate the influence of 9-AGNR surface coverage on GNR quality and orientation on the growth substrate and after substrate transfer. We characterize the overall GNR quality and the alignment of high- and low-coverage samples by scanning tunneling microscopy (STM) and polarized Raman spectroscopy. We extend a Gaussian distribution model to extract the GNRs' angle distribution (quality of alignment) and to quantify the polarization-independent Raman signal (isotropic contribution) upon growth on vicinal template surfaces and after substrate transfer. By applying this model to different coverages of GNRs and substrates, we unveil the main parameters that influence the GNRs' quality of alignment and give rise to the isotropic contribution to the Raman signal.

**Results and Discussion**

To synthesize aligned 9-AGNRs the precursor monomer 3′,6′-di-iodine-1,1′:2′,1′′-terphenyl (DITP)[34] is deposited on a vicinal catalytic surface (Au(788)) followed by two annealing steps to activate the polymerization and cyclohydrogenation reactions[5,2,4]. Samples are prepared with two different coverages (~ 0.4 monolayer and ~1 full monolayer, ML, herein referred to as low- and high-coverage samples, respectively) as shown in Fig. 1. The vicinal surface enables the growth of GNRs along the low-coordination sites of the Au(788) step edges, which act as favorable nucleation sites[43,44]. This allows GNRs to grow gradually with deposition time, and after 8 minutes with a fixed deposition rate of 1 Å /min, a full monolayer of aligned 9-AGNRs (high-coverage sample) is formed. A representative STM image of a high-coverage sample with 9-AGNRs of an average length of 34 nm is shown in Fig. 1a, the corresponding GNR length histogram is given in Fig. S1a. For the low-coverage 9-AGNR samples, a deposition time of 3 minutes is used (with a fixed deposition rate of 1 Å /min), which provides just enough precursor molecules for individual 9-AGNRs to grow along all Au(788) step edges, resulting in an average GNR length of 37 nm (Fig. 1c, see Fig. S1b for the length histogram).

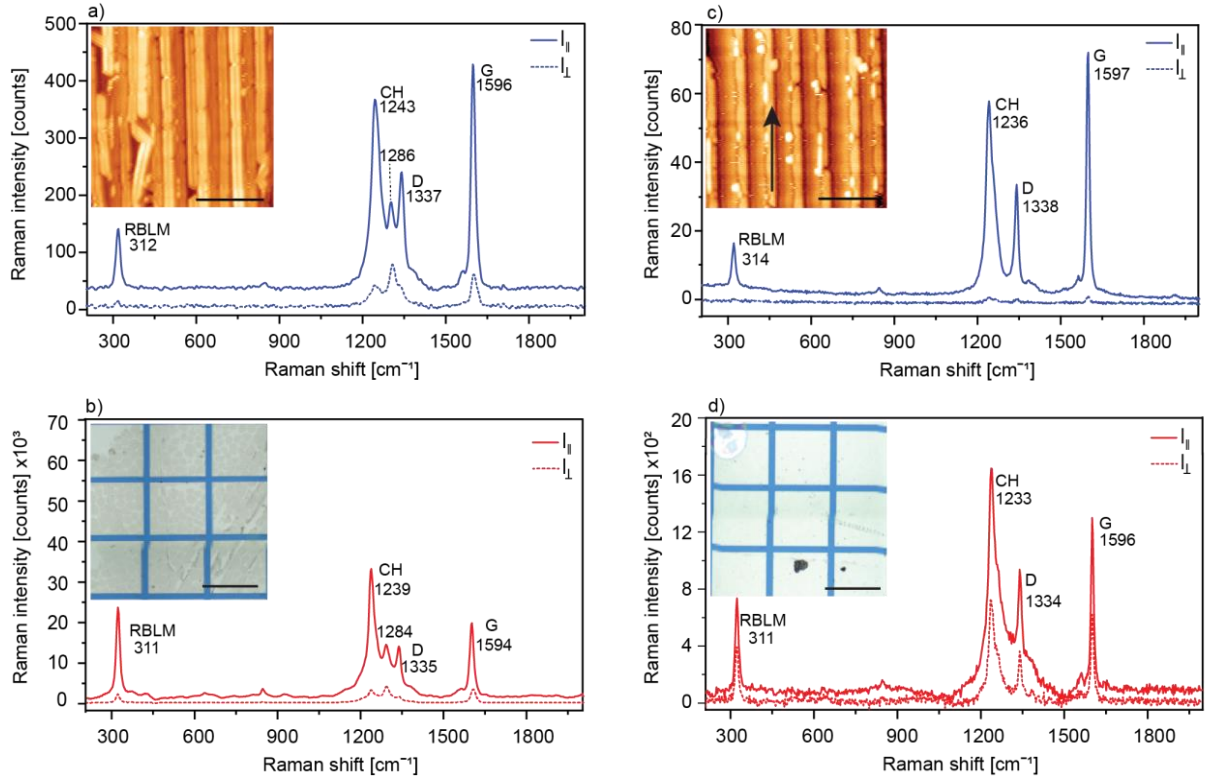

**Figure 1**: Characterization of aligned 9-AGNRs samples at high (a,b) and low (c,d) coverage before (a,c, blue spectra) and after (b,d, red spectra) substrate transfer. Raman spectra of the high-coverage sample on Au(788) (a) and after substrate transfer onto a Raman-optimized substrate (ROS)(b)[36]. The spectra are acquired with an excitation wavelength of 785 nm under vacuum conditions with polarization parallel ($I_{\parallel}$) to the GNR alignment direction (along the Au(788) step edges) (full line), and perpendicular ($I_{\perp}$) to the GNR alignment direction (dashed line). The inset in panel (a) shows a STM topography image for the high-coverage sample on Au(788), with a scale bar of 10 nm ($V_b$ =-1.5 V, $I_t$ =0.3 nA). The inset in panel (b) shows an optical micrograph of ribbons transferred onto a ROS, with a scale bar of 180 μm. Raman spectra of the low-coverage sample (c) on Au(788), and (d) after substrate transfer onto a ROS, with polarizations/full vs dashed lines as indicated above. The inset in panel (c) shows a STM image for the low-coverage sample on Au(788), with a black arrow highlighting the GNR growth direction along the Au(788) step edges of ($V_b$ =-1.5 V, $I_t$ =0.3 nA, scale bar: 10 nm). The inset in panel (d) shows an optical micrograph of GNRs transferred onto a ROS, with a scale bar of 180 μm.

While STM is a powerful technique to characterize the atomic structure of GNRs, to determine their surface coverage and local alignment on metallic growth substrates, it cannot be applied after transfer to insulating device substrates, which are normally based on SiO$_2$/Si. To follow the structural quality and alignment of GNRs on the growth substrate and upon substrate transfer, we thus carry out a detailed Raman spectroscopy investigation. Raman spectroscopy is a powerful technique for characterizing sp$^2$-hybridized carbon materials by identifying their Raman-active phonons[45–47]. The vibrational fingerprints of GNRs are named in analogy to the terminology of Raman-active phonons in other sp$^2$-hybridized carbon nanomaterials such as graphite[45], graphene[48], and CNTs[46]. The most prominent active mode in the high-frequency range of GNRs is the G mode at ~1600 cm$^{-1}$, which corresponds to the stretching of carbon-carbon bonds within the sp$^2$ lattice of the ribbon[37,49,50]. Besides the G mode, in the high-frequency region, the D and CH modes are observed[37,49,51] between 1100 and 1400 cm$^{-1}$, which are fingerprints of the GNRs' confinement-derived vibration modes and their hydrogen-passivated edges, respectively. At low frequencies, two further modes are observed: the radial-breathing-like mode (RBLM), which is related to the ribbon width [37,38,50,51], and the longitudinal compressive mode (LCM), which is a length-dependent mode universally present in AGNRs[38].

Here, Raman characterization of the high- and low-coverage samples is carried out in a home-built vacuum chamber (~$10^{-2}$ mbar) to prevent photochemical reactions during the measurements[29,36]. Additionally, an optimal mapping approach (maps of 10 µm x 10 µm) is adopted to obtain the average characteristics of 9-AGNRs with a high signal-to-noise ratio[36]. Figs. 1a and 1c show the Raman profiles for both high- and low-coverage 9-AGNR samples, respectively, acquired with a 785 nm wavelength (1.58 eV) laser on Au(788), with light polarized parallel to the nominal GNR alignment direction (along the Au(788) step edges) (full line spectrum) and perpendicular to it (dashed line spectrum). The spectra reveal the main 9-AGNR Raman active modes, namely the RBLM, CH, D, and G modes with frequencies of 312, ~1235, 1337, and 1596 cm$^{-1}$, respectively. Interestingly, for the high-coverage samples (shown in Figs. 1a and 1b) the CH mode is observed at slightly higher frequencies than that for the low-coverage samples (~1241 vs 1235 cm$^{-1}$). Another difference is the mode at ~1285 cm$^{-1}$ only resolved for the high-coverage case. These differences could indicate that at higher coverage, the nearby GNRs are more likely to be in closer proximity, possibly forming contacts or bundles, leading to higher inter-ribbon interactions. Further studies will be conducted to examine this phenomenon in more detail.

To explore and exploit GNRs' electronic properties in a device configuration, a substrate transfer step is required to transfer the GNRs from the catalytic growth surface to the target device substrate. For transferring aligned GNRs, an electrochemical delamination method is used[29,36], which was primarily developed to transfer graphene layers grown by chemical vapor deposition[52,53]. This transfer method is based on the formation of hydrogen bubbles from water electrolysis at the GNR/Au(788) interface and the use of a poly(methyl methacrylate) (PMMA) layer as polymer support upon GNR delamination. As previously demonstrated, GNRs are extremely sensitive to the electrochemical delamination transfer parameters, such as delamination time, applied current, and PMMA thickness[36]. Raman spectroscopy has previously been proven to be the method of choice to assess the quality of GNRs and monitor changes upon transfer by detecting the Raman shift, relative intensities, and peak widths of the vibrational fingerprints [37,36].

Here, we transfer both high- and low-coverage 9-AGNR samples onto Raman-optimized substrates (ROS), consisting of a Si/SiO$_2$ (285 nm) substrate with a 80 nm Au layer and a 40 nm Al$_2$O$_3$ top layer. ROS allows for signal enhancement factors of up to 120 times in comparison with standard SiO$_2$/Si[36]. Figs. 1b and 1d show the Raman profiles for both high- and low-coverage 9-AGNR samples transferred to ROS, respectively. While both samples show the presence of all Raman active modes as measured on the Au(788) growth substrate, significant differences between the high- and low-coverage samples are observed after substrate transfer. To follow the GNRs' structural quality, we first extract the full width at half maximum (FWHM) of the RBLM, D, CH, and G peaks. The average Raman profile of the high-coverage sample shows similar FWHMs before and after the substrate transfer (~14 cm$^{-1}$ for RBLM, ~24 cm$^{-1}$ for CH, ~18 cm$^{-1}$ for D, and ~12 cm$^{-1}$ for G mode). On the other hand, the low-coverage sample shows significant broadening upon transfer for RBLM (from ~12 to 19 cm$^{-1}$), CH (from ~20 to 40 cm$^{-1}$), D (from ~11 to 25 cm$^{-1}$), and G modes (from ~10 to 18 cm$^{-1}$), accompanied by an overall decrease in the signal-to-noise ratio, indicating that the GNRs' overall quality and quantity are not entirely preserved. This behavior suggests that 9-AGNRs growing along the Au(788) step edges have a stronger physical interaction with the gold substrate, and are less likely to transfer efficiently and without defects.

Besides monitoring the GNRs' structural quality, Raman spectroscopy is also a powerful technique to characterize their orientation due to GNR's anisotropy. Raman polarization anisotropy ($P$) is the most used parameter for assessing GNR's average orientation. We extract $P$ for all Raman active modes as a function of coverage before and after substrate transfer. Fig. 1 shows $I_\parallel$ (full spectra) and $I_\perp$ (dashed spectra), which represent Raman spectra for incoming and scattered light polarized parallel and perpendicular to the GNRs' nominal alignment direction (the Au(788) step direction), respectively, for both high- and low-coverage 9-AGNR samples on the growth surface (Figs. 1a and 1c, respectively) and on the ROS (Figs. 1b and 1d, respectively). The high-coverage sample shows an anisotropy of $P = 0.86$ on the Au(788) substrate, which only slightly decreases to $P = 0.85$ after substrate transfer, indicating that the degree of GNR alignment is largely preserved. Similar preservation of GNR alignment upon substrate transfer was reported previously for high-coverage 7-AGNRs and 9-AGNRs[29,27]. On the other hand, the low-coverage sample shows very different behavior, with polarization anisotropy decreasing significantly from $P = 0.95$ on Au (788) to $P = 0.58$ upon substrate transfer.

The preservation of the overall degree of alignment is thus clearly coverage-dependent. In a high-coverage sample, the 9-AGNR layer seems to behave very much like a film, with very low GNR mobility during substrate transfer, which preserves the overall degree of GNR alignment. This film-like behavior is absent in the low-coverage 9-AGNR samples, where only individual GNRs grow along the Au(788) step edges. In addition, GNRs growing solely along the step edge appear to show a stronger interaction with the substrate (due to the higher site reactivity), contributing to a less efficient transfer. Although the polarization anisotropy provides clear information on the overall degree of GNR alignment, it does not include detailed information on the angular distribution nor the isotropic contributions of small polyaromatic hydrocarbons (PAHs), short GNRs, or PMMA residues to the overall Raman intensity. To address that, we model the GNRs' angular distribution as a Gaussian distribution and quantify the isotropic contribution to the Raman intensity by taking into account a polarization angle-independent intensity in addition to the usual polarization-dependent intensity distribution resulting from the aligned GNRs.

Raman profiles are obtained by polarizing the incoming and scattered light in parallel ("VV configuration") with different angles between the nominal GNR alignment direction and the polarization of the incident light. Using the VV configuration implies that for the Raman resonant modes, the intensity of the GNR modes is projected to be $cos^4(\vartheta)$ polarization-dependent, which results from a product of two

$cos^2(\vartheta)$ factors, one for photon absorption and the other for photon emission, Eq. (1)[54–59]. This means that the Raman signal is maximum with the incident polarization parallel to the ribbon axis (0°, 180°) and zero when perpendicular to it (90°, 270°) (Fig. 2). In Eq. (1), $\vartheta_0$ is the orientation of the long axis of the GNR with respect to an arbitrary in-plane axis, and $\vartheta$ is the direction of the light polarization. Due to the significant absorption anisotropy of the quasi-1D GNRs, all Raman modes exhibit roughly the same polarization dependency.

$$I_{Raman}^{pol}(\vartheta) \approx cos^4(\vartheta - \vartheta_0) \qquad (1)$$

To model the polarized Raman intensity as a function of polarization direction for samples with many GNRs that are not perfectly aligned, we assume that the GNRs are on average aligned along the direction $\vartheta_0$, with a normalized Gaussian distribution of angles $\vartheta$ [29,57], $G(\vartheta)$ as in Eq. (2), where $\sigma$ is the standard deviation that is related to the FWHM=$2\sigma\sqrt{2 \ln 2}$.

$$G(\vartheta) = \frac{1}{\sigma\sqrt{2\pi}} e^{-\frac{(\vartheta - \vartheta_0)^2}{2\sigma^2}} \qquad (2)$$

In addition to the GNRs aligned according to $G(\vartheta)$, we assume that small PAHs, short GNRs, or polymer residues from the substrate transfer step give rise to an isotropic Raman intensity contribution that does not depend on the polarization direction. Such polarization-independent Raman intensity contributions have been observed in the case of CNTs, owing to the presence of amorphous carbon and/or carbon nanocomposites[54,57,59,60]. To account for such contributions in the Raman and optical absorption intensities in the single-walled carbon nanotubes (SWCNTs) samples, an angle-independent component was introduced to the fit function[57]. Here, we use a similar strategy to account for the isotropic contribution of such Raman polarization-independent intensity (i.e. intensity that does not depend on $\vartheta$) by adding a constant $H$ (Eq. (2)).

$$H(\vartheta) = H \qquad (3)$$

The total angular distribution function $D(\vartheta)$ of species contributing to the Raman intensity (GNRs, PAHs, PMMA residues) is then defined in Eq. (4) as the sum of $G(\vartheta)$ weighted with the fraction of the surface area $A$ exhibiting aligned GNRs and of $H(\vartheta)$ weighted with the fraction of the surface area $B$ producing the isotropic contribution.

$$D(\vartheta) = A \cdot G(\vartheta) + B \cdot H \qquad (4)$$

For $A$ and $B$ to be meaningful fit parameters, it is necessary to properly normalize the constant $H$ and the Gaussian distribution $G(\vartheta)$. The Gaussian distribution $G(\vartheta)$ is normalized to 1 (integral from 0° to 360° is 1), thus we normalize $H(\vartheta)$ correspondingly to an integral from 0° to 360° of 1, implying $H=1/360$. Eq. (5) gives the resulting normalized angular distribution function $D(\vartheta)$, including the homogeneous background.

$$D(\vartheta) = \frac{A}{\sigma\sqrt{2\pi}} e^{-\frac{(\vartheta-\vartheta_0)^2}{2\sigma^2}} + \frac{B}{360} \tag{5}$$

The expected Raman signal $I_{exp}(\vartheta)$ is obtained from the convolution of $I_{Raman}^{pol}(\vartheta)$ and $D(\vartheta)$:

$$I_{exp}(\vartheta) = \int_{0°}^{360°} I_{Raman}^{pol}(\varphi) \cdot D(\vartheta - \varphi) d\varphi \tag{6}$$

$$= \int_{0°}^{360°} \cos^4(\varphi) \cdot \left( \frac{A}{\sigma\sqrt{2\pi}} e^{-\frac{(\vartheta-\varphi-\vartheta_0)^2}{2\sigma^2}} + \frac{B}{360} \right) d\varphi \tag{7}$$

$$= A \cdot \int_{0°}^{360°} \cos^4(\varphi) \cdot \frac{1}{\sigma\sqrt{2\pi}} e^{-\frac{(\vartheta-\varphi-\vartheta_0)^2}{2\sigma^2}} d\varphi + \frac{B}{360} \cdot \int_{0°}^{360°} \cos^4(\varphi) \, d\varphi \tag{8}$$

The first integral in Eq. (8) is simply the convolution of $\cos^4(\vartheta)$ with a normalized Gaussian distribution, whereas the second integral is a constant that equals $3 \cdot 180°/4$, in degrees.

From that, we obtain:

$$I_{exp}(\vartheta) = A \cdot \int_{0°}^{360°} \cos^4(\varphi) \cdot \frac{1}{\sigma\sqrt{2\pi}} e^{-\frac{(\vartheta-\varphi-\vartheta_0)^2}{2\sigma^2}} d\varphi + \frac{B}{360} \cdot \frac{3 \cdot 180}{4}$$

$$= A \cdot \int_{0°}^{360°} \cos^4(\varphi) \cdot \frac{1}{\sigma\sqrt{2\pi}} e^{-\frac{(\vartheta-\varphi-\vartheta_0)^2}{2\sigma^2}} d\varphi + B \cdot \frac{3}{8} \tag{9}$$

In Fig. 2, we illustrate the main contributions to the expected Raman intensity: The $\cos^4(\vartheta)$ polarization-dependent Raman intensity (in red) and the angular distribution function $D(\vartheta)$ (in blue) which includes aligned GNRs with an angle distribution $G(\vartheta)$ and the isotropic contribution (polarization-independent component) to account for the homogeneous background.

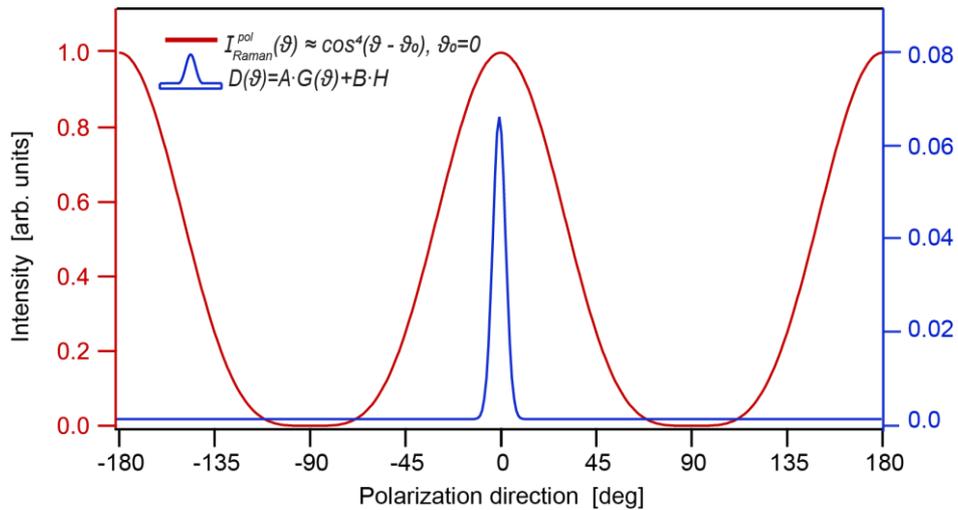

Figure 2: Illustration of the major contributions to the expected polarized Raman intensity as described by Eq. (9). The Raman intensity of a single GNR follows $\cos^4(\vartheta)$ dependence (red). In blue, the angular distribution function $D(\vartheta)$ is shown, which includes the normalized Gaussian distribution of angles and the normalized isotropic contribution. For this particular plot, $\sigma=3°$, $A=0.5$, and $B=0.5$ have been used.

By using Eq. (9) for the fitting of angle-dependent polarized Raman intensity data, it is possible to extract the following relevant information: $\vartheta_0$, the azimuthal angle along which GNRs are preferentially aligned (the center of the Gaussian distribution); the fraction $A$ of surface area exhibiting aligned GNRs; $\sigma$, the width of the Gaussian distribution characterizing the angular distribution around $\vartheta_0$ (here defined as the quality of alignment); and the fraction $B$ of the surface area contributing to the isotropic, polarization-independent Raman signal. From $A$ and $B$, we can define the "overall disorder" present on the surface ($OD$) as follows (Eq. (10)):

$$OD = \frac{B}{(B+A)} \cdot 100\% \qquad (10)$$

To illustrate the effect of increasing $OD$ for specific values of $\sigma$ (and vice versa) on the Raman intensity, we plot the Raman intensity versus polarization angle for varying $\sigma$ from 1° to 30° and $OD$ from 0% up to 30%, as shown in Fig. S2. When keeping $\sigma$ constant and increasing $OD$ from 0% to 30%, we observe an increase in baseline with a higher vertical offset of Raman intensity for larger values of $OD$. The increase in the baseline is a direct indication of the increased disorder present on the surface. On the other hand, when we increase the width of $\sigma$ from 1° to 30° while keeping $OD$ constant, we observe a significant decrease in Raman intensity along with significant broadening – a direct measure of how well the GNRs are aligned and their angle distribution within the sample. Furthermore, we discuss the impact of both $\sigma$ and $OD$ on the $P$ in the supplementary information (Fig. S3). We observe that $P$ is significantly affected by $OD$, with an almost linear decrease as $OD$ increases, whereas for $\sigma$, $P$ is only significantly affected for $\sigma > 15°$. This indicates that $P$ is a reasonably good indicator of the combined impact of $\sigma$ and $OD$ for larger values of $\sigma$, but rather insensitive to width differences for narrow ($\sigma <$ 15°) angle distributions.

To investigate the influence of GNR coverage and substrate transfer on $\sigma$ and $OD$, we fit all Raman active modes of high- and low-coverage 9-AGNR samples on both the growth and ROS using Eq. (8). Fig. 3 shows the G mode peak intensity as a function of the polarization angle $\vartheta$ for the VV configuration and the related polar diagrams for both high- (Fig. 3a and 3b) and low-coverage samples (Fig. 3c and 3d) on Au(788) (in blue) and after substrate transfer (in red), respectively (see Figs. S4 and S5 for similar plots for CH, D, and RBLM modes). The intensity of the G mode as a function of polarization angle (-90° to +90°) is determined from Raman maps of 10 x 10 pixels in vacuum conditions using a 785 nm laser energy. We fit the polar dependence of the Raman intensity using Eq. (9).

It is important to note that the G peak of GNRs is composed of two in-plane optical modes: a transverse-optical (TO) and a longitudinal-optical (LO) mode[2,37,49]. The longitudinal optical (LO) (G1 peak, phonon mode with $A_g$ symmetry) generates an atomic displacement parallel to the GNR axis and has a maximum intensity along the GNR, whereas the transverse optical (TO) (G2 peak, phonon mode with $B_{1g}$ symmetry) has a maximum intensity perpendicular to the GNR axis. Here, all the measurements are done using a 785 nm laser energy, and in our spectra, we only resolve the LO (G1) mode, which for simplicity we call G mode.

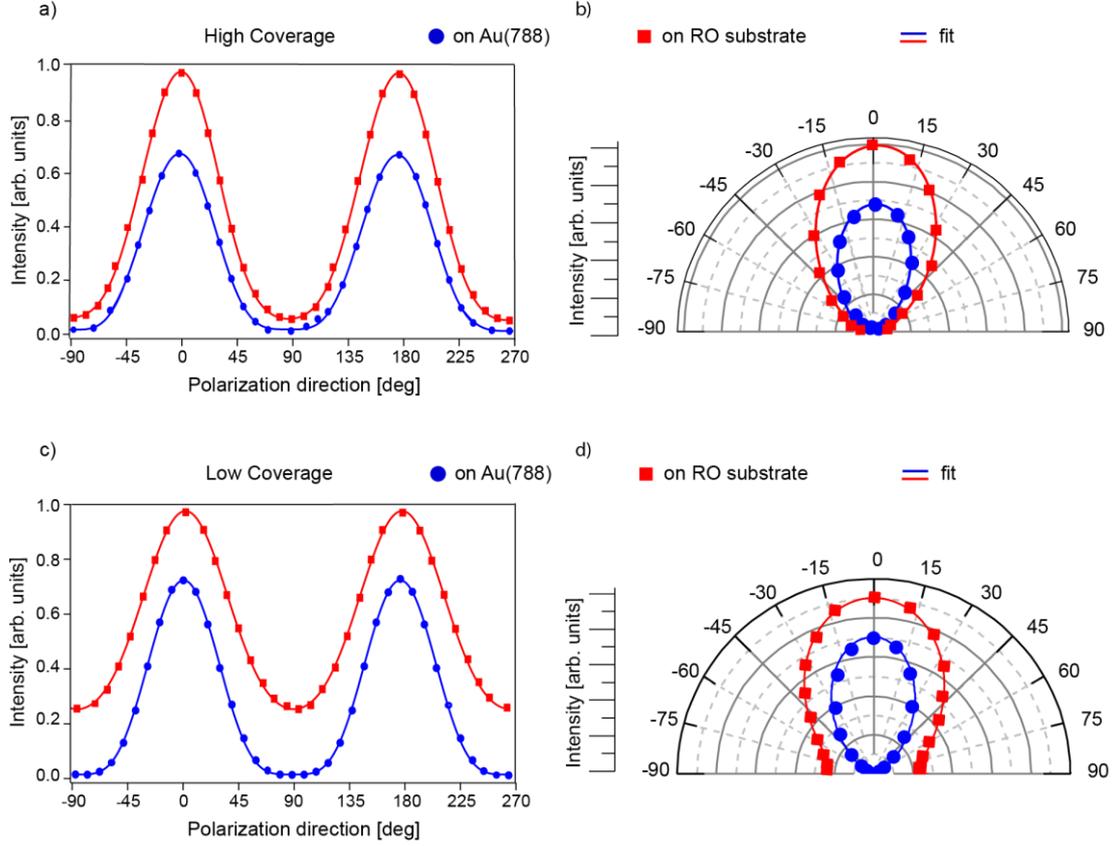

Figure 3: Polarized Raman intensity of G mode (785 nm, VV configuration). (a, c) G mode intensity as a function of polarization angle $\vartheta$ for high- and low-coverage samples on Au(788) (blue circles) and after substrate transfer onto ROS (red squares). Blue and red solid lines represent data fits using Eq. (9). (b, d) Polar diagrams showing G mode intensities for high- and low-coverage samples on Au(788) (blue circles) and after transfer to ROS (red squares). Blue and red solid lines represent fits to the measured data using Eq. (9).

Table 1 shows the relevant fitting results for the G mode data: the quality of alignment ($\sigma$), and the overall disorder on the surface (*OD*) for both high- and low-coverage samples on Au(788) and on the ROS extracted from Fig. 3. For comparison, we also show the values of Raman polarization anisotropy for all cases; see Table S1 and Table S2 for *P*, $\sigma$, and *OD* values for RBLM, CH, and D modes for both high- and low-coverage samples on Au(788) and the ROS.

|  | High-coverage sample | | Low-coverage sample | |
| --- | --- | --- | --- | --- |
|  | Au(788) | ROS | Au(788) | ROS |
| Quality of alignment ($\sigma$) [°] | 3 ± 1 | 13 ± 1 | 1.0 ± 0.1 | 22 ± 3 |
| Overall disorder on the surface (*OD*) [%] | 8 ± 1 | 13 ± 1 | 8 ± 3 | 39 ± 3 |
| Raman polarization anisotropy (*P*) | 0.86 | 0.85 | 0.95 | 0.58 |

Table 1: Comparison of high- and low-coverage samples ($\sigma$, *OD*, and *P*) on Au(788) and after substrate transfer onto ROS.

A small value of $\sigma$ means a narrow-angle distribution and, thus, a high degree/ quality of uniaxial alignment. We observe that $\sigma$ increases upon substrate transfer from 3° to 13° for the high-coverage sample and from 1° to 22° for the low-coverage one, indicating poorer quality of alignment after the transfer, especially for the low-coverage sample. This behavior can be attributed to two main factors: the strong interaction of GNRs with the Au(788) step edges, making it less likely for the GNRs to transfer efficiently, and the increased GNR mobility (especially for the low surface coverage), which increases the angle distribution within GNRs upon transfer. In addition, when comparing $\sigma$ for both coverages of 9-AGNRs on Au(788), we observe slightly lower $\sigma$ values for the low-coverage sample, indicating a better quality of alignment for 9-AGNRs grown only on the step edges of the Au(788). This is also reflected in the polarization anisotropy measurements, with $P = 0.95$ for the low-coverage 9-AGNR sample and $P = 0.85$ for the high-coverage one on Au(788), indicating the higher degree of alignment of 9-AGNRs grown along the step-edge compared to the complete monolayer (see Fig. S6 for a STM image of a 9-AGNR high-coverage sample highlighting the presence of smaller GNRs growing perpendicular to the terraces in some areas).

From Table 1, it can also be seen that substrate transfer increases the $OD$ fraction from 8 to 13% and from 8 to 39% for high- and low-coverage samples, respectively. The significant increase for the low-coverage samples is explained by GNRs' strong interaction with the Au(788) step edges which makes the low-coverage samples much higher susceptibility to inefficient transfer, leading to a higher prevalence of partially broken GNRs. In addition, the low-coverage sample leaves considerably more exposed gold substrate to PMMA and other impurities that might react with the Au surface and transfer along with the GNRs, increasing the $OD$ on the transferred surface.

On the Au(788) growth substrate we observe a similar $OD$ for both high- and low-coverage samples (8%). The disorder observed on the growth substrate may originate from short (and thus non-aligned) GNRs, irregularly fused precursor monomers, or also the presence of impurities from the precursor monomer.

**Conclusions**

In this study, we employed polarized Raman spectroscopy and scanning tunneling microscopy to characterize and quantify the structural quality and degree of alignment 9-AGNRs in samples with different surface coverages on both their growth substrate and after substrate transfer. Using an extended data analysis model, which describes GNR alignment by a Gaussian distribution of angles, allowed us to extract both the quality of alignment ($\sigma$) and the overall surface disorder ($OD$).

Our results show that low-coverage samples exhibit better uniaxial alignment than high-coverage samples on the growth substrate. This behavior results from GNRs in low-coverage samples growing preferentially along the step-edges of Au(788), as observed in our STM investigations. However, upon transfer, the quality of alignment of low-coverage samples is significantly reduced, which we attribute mostly to the strong interaction of GNRs with the Au(788) step edges as well as increased GNR mobility, whereas high-coverage samples show better alignment preservation upon substrate transfer, owing to the densely packed GNR film facilitating the transfer process. With the extended model developed in this study, we also quantified the *OD*, which results in an isotropic (polarization-independent) contribution to the Raman intensity. After substrate transfer, low-coverage samples show systematically higher *OD* values than high-coverage samples (39% vs 13% respectively). The significantly higher *OD* for low-coverage samples is associated with the strong interaction of GNRs to the Au(788) step edges, making it less likely for the GNRs to transfer efficiently, as well as to the fact that more gold surface area is exposed to PMMA and other impurities that may react with the metal and transfer along with the GNRs to the target substrate.

Overall, our results shed light on the crucial role of surface coverage in determining the degree of alignment and the *OD* present on the surface on both the Au(788) growth surface and, in particular, after substrate transfer. Our extended model provided a quantitative description of GNR alignment and quality, which is a pivotal step toward the development of integrated GNR-based nanoelectronic devices and establishes polarized Raman as the method of choice for tracking GNR quality and degree of alignment during transfer and device fabrication steps.

**Methods**

*On-Surface Synthesis and STM Characterization of 9-AGNRs*

The Au(788) single crystal growth substrate (MaTecK GmbH, Germany) was cleaned in ultra-high vacuum (UHV) with two cycles of sputtering at 1 kV $Ar^+$ for 10 minutes and annealing at 420 °C for 10 minutes. The 9-AGNR precursor monomer 3′,6′-di-iodine-1,1′:2′,1″-terphenyl (DITP) was then sublimated onto the clean Au surface from a quartz crucible heated to 70 °C while the substrate remained at room temperature[34]. A quartz microbalance was used to control the deposition rate of the precursor molecules at 1 Å /min. The deposition rate is not calibrated to correspond to the true surface coverage, but only give a relative measurement that is then calibrated by STM. High- and low-coverage samples were obtained by DITP deposition for 8 and 3 minutes, respectively. Following deposition, the substrate was heated to 200 °C (0.5 K/s) for 10 minutes to initiate DITP polymerization, followed by annealing at 400 °C (0.5 K/s) for 10 minutes to form the GNRs by cyclodehydrogenation [5,2,4,34].

Scanning tunneling microscopy images of 9-AGNRs grown on Au(788) were acquired at room temperature using a Scienta Omicron VT-STM. Topographic images were acquired in constant current mode using a sample bias of -1.5 V and a setpoint current of 0.03 nA.

*Substrate transfer of 9-AGNRs*

Transfer of 9-AGNRs from their Au(788) growth substrate to the Raman-optimized substrates (ROS) was done by electrochemical delamination transfer[29,36]. First, a support layer of poly(methyl methacrylate) (PMMA) was spin-coated (4 PMMA layers, 2500 rpm for 90 s) on the 9-AGNR/Au(788) samples, followed by a 10-minute curing process at 80 °C. To shorten the time required for PMMA delamination, PMMA was removed from the Au(788) crystal's edges using a 80-minute UV exposure, followed by a 3-minute development in water/isopropanol. Electrochemical delamination was performed in an aqueous solution of NaOH (1 M) as the electrolyte. A DC voltage of 5V (current ~0.2 A) was applied between the PMMA/9-AGNR/Au(788) cathode and a glassy carbon electrode used as the anode. At the interface between PMMA/GNRs and Au, hydrogen bubbles form, resulting in the delamination of the PMMA/GNR layer from the Au(788) surface. The delaminated PMMA/GNR layer was left in ultra-pure water for 5 minutes before being transferred to the target substrate. To increase the adhesion between the target substrate and the PMMA/GNR layer, the sample was annealed for 10 minutes at 80°C and then 20 minutes at 110°C. Finally, the PMMA was dissolved in acetone for 15 minutes, and the resulting GNR/ROS was washed with ethanol and ultrapure water.

*Raman Spectroscopy*

Raman spectroscopy measurements were obtained using a WITec confocal Raman microscope (WITec Alpha 300R) with a laser line of 785 nm (1.5 eV) and a power of 40 mW. A 50× microscope objective (0.55 numerical aperture) with a working distance of 9.1 mm (resulting laser spot size of 600 nm) was used to focus the laser beam onto the sample and collect the scattered light. Calibration of Raman spectra was performed using the Si peak at 520.5 cm$^{-1}$. Also, the laser wavelength, power, and integration time were optimized for each substrate to maximize signal while minimizing sample damage. Furthermore, to avoid sample damage, a Raman mapping approach with 10×10 pixels (10×10 μm) was used and samples were measured in a home-built vacuum suitcase with pressure ~10$^{-2}$ mbar. The vacuum chamber was mounted on a piezo stage for scanning.

The "VV" configuration was used for polarized Raman measurements, with the polarizer oriented parallel to the polarization of the incident light. A motorized half-wave plate was used to change the polarization direction of the incident laser beam from -90° to +90° in steps of 10°. To control the scattered light direction and keep it parallel in the detection path a manual analyzing polarizer and a λ/2 plate were inserted before the detector. For measurements with 785 nm excitation wavelength, the scattered signal was detected with an analyzing polarizer coupled with a 300 mm lens-based spectrometer with a grating of 300 g mm$^{-1}$ (grooves/mm) and equipped with a cooled deep-depletion CCD.

*Raman data processing*

Using the WITec software, a cosmic ray filter was applied to all raw maps for removing signatures of photoluminescence. Afterward, the Raman maps were averaged and polynomial background subtraction was applied, followed by batched fitting with a Lorentzian function for all polarization angles between -90° to 90° for each Raman mode. The fitting using Eq. (9) was done in IGOR Pro software (Wavemetrics Inc.), and the fitting parameters were obtained through the lowest stable Chi-square values.

**Acknowledgments**

This work was supported by the Swiss National Science Foundation under grant no. 200020_182015, the European Union Horizon 2020 research and innovation program under grant agreement no. 881603 (GrapheneFlagship Core 3), and the Office of Naval Research BRC Program under the grant N00014-

18-1-2708. The authors also greatly appreciate the financial support from the Werner Siemens Foundation (CarboQuant). R.D acknowledges funding from the University of Bern.

*For the purpose of Open Access, the author has applied a CC BY public copyright license to any Author Accepted Manuscript version arising from this submission.*

**REFERENCES:**

[1] L. Yang, C.-H. Park, Y.-W. Son, M.L. Cohen, S.G. Louie, Quasiparticle Energies and Band Gaps in Graphene Nanoribbons, Phys. Rev. Lett. 99 (2007). https://doi.org/10.1103/PhysRevLett.99.186801.

[2] L. Talirz, H. Söde, T. Dumslaff, S. Wang, J.R. Sanchez-Valencia, J. Liu, P. Shinde, C.A. Pignedoli, L. Liang, V. Meunier, N.C. Plumb, M. Shi, X. Feng, A. Narita, K. Müllen, R. Fasel, P. Ruffieux, On-Surface Synthesis and Characterization of 9-Atom Wide Armchair Graphene Nanoribbons, ACS Nano. 11 (2017) 1380–1388. https://doi.org/10.1021/acsnano.6b06405.

[3] A. Kimouche, M.M. Ervasti, R. Drost, S. Halonen, A. Harju, P.M. Joensuu, J. Sainio, P. Liljeroth, Ultra-narrow metallic armchair graphene nanoribbons, Nat. Commun. 6 (2015) 10177. https://doi.org/10.1038/ncomms10177.

[4] L. Talirz, P. Ruffieux, R. Fasel, On-Surface Synthesis of Atomically Precise Graphene Nanoribbons, Adv Mater. 28 (2016) 6222–6231. https://doi.org/10.1002/adma.201505738.

[5] J. Cai, P. Ruffieux, R. Jaafar, M. Bieri, T. Braun, S. Blankenburg, M. Muoth, A.P. Seitsonen, M. Saleh, X. Feng, K. Müllen, R. Fasel, Atomically precise bottom-up fabrication of graphene nanoribbons, Nature. 466 (2010) 470–473. https://doi.org/10.1038/nature09211.

[6] O. Gröning, S. Wang, X. Yao, C.A. Pignedoli, G. Borin Barin, C. Daniels, A. Cupo, V. Meunier, X. Feng, A. Narita, K. Müllen, P. Ruffieux, R. Fasel, Engineering of robust topological quantum phases in graphene nanoribbons, Nature. 560 (2018) 209–213. https://doi.org/10.1038/s41586-018-0375-9.

[7] D.J. Rizzo, G. Veber, T. Cao, C. Bronner, T. Chen, F. Zhao, H. Rodriguez, S.G. Louie, M.F. Crommie, F.R. Fischer, Topological band engineering of graphene nanoribbons, Nature. 560 (2018) 204–208. https://doi.org/10.1038/s41586-018-0376-8.

[8] Q. Sun, O. Gröning, J. Overbeck, O. Braun, M.L. Perrin, G.B. Barin, M.E. Abbassi, K. Eimre, E. Ditler, C. Daniels, V. Meunier, C.A. Pignedoli, M. Calame, R. Fasel, P. Ruffieux, Massive Dirac Fermion Behavior in a Low Bandgap Graphene Nanoribbon Near a Topological Phase Boundary, Adv. Mater. 32 (2020) 1906054. https://doi.org/10.1002/adma.201906054.

[9] T. Cao, F. Zhao, S.G. Louie, Topological Phases in Graphene Nanoribbons: Junction States, Spin Centers, and Quantum Spin Chains, Phys. Rev. Lett. 119 (2017) 076401. https://doi.org/10.1103/PhysRevLett.119.076401.

[10] P. Ruffieux, S. Wang, B. Yang, C. Sánchez-Sánchez, J. Liu, T. Dienel, L. Talirz, P. Shinde, C.A. Pignedoli, D. Passerone, T. Dumslaff, X. Feng, K. Müllen, R. Fasel, On-surface synthesis of graphene nanoribbons with zigzag edge topology, Nature. 531 (2016) 489–492. https://doi.org/10.1038/nature17151.

[11] J.P. Llinas, A. Fairbrother, G. Borin Barin, W. Shi, K. Lee, S. Wu, B. Yong Choi, R. Braganza, J. Lear, N. Kau, W. Choi, C. Chen, Z. Pedramrazi, T. Dumslaff, A. Narita, X. Feng, K. Müllen, F. Fischer, A. Zettl, P. Ruffieux, E. Yablonovitch, M. Crommie, R. Fasel, J. Bokor, Short-channel field-effect transistors with 9-atom and 13-atom wide graphene nanoribbons, Nat. Commun. 8 (2017) 633. https://doi.org/10.1038/s41467-017-00734-x.

[12] M. El Abbassi, M.L. Perrin, G.B. Barin, S. Sangtarash, J. Overbeck, O. Braun, C.J. Lambert, Q. Sun, T. Prechtl, A. Narita, K. Müllen, P. Ruffieux, H. Sadeghi, R. Fasel, M. Calame, Controlled Quantum Dot

Formation in Atomically Engineered Graphene Nanoribbon Field-Effect Transistors, ACS Nano. 14 (2020) 5754–5762. https://doi.org/10.1021/acsnano.0c00604.

[13] O. Braun, J. Overbeck, M. El Abbassi, S. Käser, R. Furrer, A. Olziersky, A. Flasby, G. Borin Barin, Q. Sun, R. Darawish, K. Müllen, P. Ruffieux, R. Fasel, I. Shorubalko, M.L. Perrin, M. Calame, Optimized graphene electrodes for contacting graphene nanoribbons, Carbon. 184 (2021) 331–339. https://doi.org/10.1016/j.carbon.2021.08.001.

[14] P.B. Bennett, Z. Pedramrazi, A. Madani, Y.-C. Chen, D.G. de Oteyza, C. Chen, F.R. Fischer, M.F. Crommie, J. Bokor, Bottom-up graphene nanoribbon field-effect transistors, Appl. Phys. Lett. 103 (2013) 253114. https://doi.org/10.1063/1.4855116.

[15] M. Ohtomo, Y. Sekine, H. Hibino, H. Yamamoto, Graphene nanoribbon field-effect transistors fabricated by etchant-free transfer from Au(788), Appl. Phys. Lett. 112 (2018) 021602. https://doi.org/10.1063/1.5006984.

[16] V. Passi, A. Gahoi, B.V. Senkovskiy, D. Haberer, F.R. Fischer, A. Grüneis, M.C. Lemme, Field-Effect Transistors Based on Networks of Highly Aligned, Chemically Synthesized $N = 7$ Armchair Graphene Nanoribbons, ACS Appl. Mater. Interfaces. 10 (2018) 9900–9903. https://doi.org/10.1021/acsami.8b01116.

[17] Z. Chen, A. Narita, K. Müllen, Graphene Nanoribbons: On-Surface Synthesis and Integration into Electronic Devices, Adv. Mater. 32 (2020) 2001893. https://doi.org/10.1002/adma.202001893.

[18] Z. Mutlu, Y. Lin, G.B. Barin, Z. Zhang, G. Pitner, S. Wang, R. Darawish, M.D. Giovannantonio, H. Wang, J. Cai, M. Passlack, C.H. Diaz, A. Narita, K. Müllen, F.R. Fischer, P. Bandaru, A.C. Kummel, P. Ruffieux, R. Fasel, J. Bokor, Short-Channel Double-Gate FETs with Atomically Precise Graphene Nanoribbons, in: 2021 IEEE Int. Electron Devices Meet. IEDM, 2021: p. 37.4.1-37.4.4. https://doi.org/10.1109/IEDM19574.2021.9720620.

[19] H. Wang, H.S. Wang, C. Ma, L. Chen, C. Jiang, C. Chen, X. Xie, A.-P. Li, X. Wang, Graphene nanoribbons for quantum electronics, Nat. Rev. Phys. 3 (2021) 791–802. https://doi.org/10.1038/s42254-021-00370-x.

[20] Y.C. Lin, Z. Mutlu, G. Borin Barin, Y. Hong, J.P. Llinas, A. Narita, H. Singh, K. Müllen, P. Ruffieux, R. Fasel, J. Bokor, Scaling and statistics of bottom-up synthesized armchair graphene nanoribbon transistors, Carbon. 205 (2023) 519–526. https://doi.org/10.1016/j.carbon.2023.01.054.

[21] J. Zhang, O. Braun, G.B. Barin, S. Sangtarash, J. Overbeck, R. Darawish, M. Stiefel, R. Furrer, A. Olziersky, K. Müllen, I. Shorubalko, A.H.S. Daaoub, P. Ruffieux, R. Fasel, H. Sadeghi, M.L. Perrin, M. Calame, Tunable Quantum Dots from Atomically Precise Graphene Nanoribbons Using a Multi-Gate Architecture, Adv. Electron. Mater. 9 (2023) 2201204. https://doi.org/10.1002/aelm.202201204.

[22] J. Zhang, L. Qian, G.B. Barin, A.H.S. Daaoub, P. Chen, K. Müllen, S. Sangtarash, P. Ruffieux, R. Fasel, H. Sadeghi, J. Zhang, M. Calame, M.L. Perrin, Ultimately-scaled electrodes for contacting individual atomically-precise graphene nanoribbons, (2022). https://doi.org/10.48550/arXiv.2209.04353.

[23] W. Niu, S. Sopp, A. Lodi, A. Gee, F. Kong, T. Pei, P. Gehring, J. Nägele, C.S. Lau, J. Ma, J. Liu, A. Narita, J. Mol, M. Burghard, K. Müllen, Y. Mai, X. Feng, L. Bogani, Exceptionally clean single-electron transistors from solutions of molecular graphene nanoribbons, Nat. Mater. 22 (2023) 180–185. https://doi.org/10.1038/s41563-022-01460-6.

[24] M. Slota, A. Keerthi, W.K. Myers, E. Tretyakov, M. Baumgarten, A. Ardavan, H. Sadeghi, C.J. Lambert, A. Narita, K. Müllen, L. Bogani, Magnetic edge states and coherent manipulation of graphene nanoribbons, Nature. 557 (2018) 691–695. https://doi.org/10.1038/s41586-018-0154-7.

[25] R.E. Blackwell, F. Zhao, E. Brooks, J. Zhu, I. Piskun, S. Wang, A. Delgado, Y.-L. Lee, S.G. Louie, F.R. Fischer, Spin splitting of dopant edge state in magnetic zigzag graphene nanoribbons, Nature. 600 (2021) 647–652. https://doi.org/10.1038/s41586-021-04201-y.

[26] E. Kan, Z. Li, J. Yang, J.G. Hou, Half-Metallicity in Edge-Modified Zigzag Graphene Nanoribbons, J. Am. Chem. Soc. 130 (2008) 4224–4225. https://doi.org/10.1021/ja710407t.

[27] S. Zhao, G.B. Barin, T. Cao, J. Overbeck, R. Darawish, T. Lyu, S. Drapcho, S. Wang, T. Dumslaff, A. Narita, M. Calame, K. Müllen, S.G. Louie, P. Ruffieux, R. Fasel, F. Wang, Optical Imaging and Spectroscopy of Atomically Precise Armchair Graphene Nanoribbons, Nano Lett. 20 (2020) 1124–1130. https://doi.org/10.1021/acs.nanolett.9b04497.

[28] S. Zhao, G.B. Barin, L. Rondin, C. Raynaud, A. Fairbrother, T. Dumslaff, S. Campidelli, K. Müllen, A. Narita, C. Voisin, P. Ruffieux, R. Fasel, J.-S. Lauret, Optical Investigation of On-Surface Synthesized Armchair Graphene Nanoribbons, Phys. Status Solidi B. 254 (2017) 1700223. https://doi.org/10.1002/pssb.201700223.

[29] B.V. Senkovskiy, M. Pfeiffer, S.K. Alavi, A. Bliesener, J. Zhu, S. Michel, A.V. Fedorov, R. German, D. Hertel, D. Haberer, L. Petaccia, F.R. Fischer, K. Meerholz, P.H.M. van Loosdrecht, K. Lindfors, A. Grüneis, Making Graphene Nanoribbons Photoluminescent, Nano Lett. 17 (2017) 4029–4037. https://doi.org/10.1021/acs.nanolett.7b00147.

[30] R. Denk, M. Hohage, P. Zeppenfeld, J. Cai, C.A. Pignedoli, H. Söde, R. Fasel, X. Feng, K. Müllen, S. Wang, D. Prezzi, A. Ferretti, A. Ruini, E. Molinari, P. Ruffieux, Exciton-dominated optical response of ultra-narrow graphene nanoribbons, Nat. Commun. 5 (2014) 4253. https://doi.org/10.1038/ncomms5253.

[31] C. Cocchi, D. Prezzi, A. Ruini, E. Benassi, M.J. Caldas, S. Corni, E. Molinari, Optical Excitations and Field Enhancement in Short Graphene Nanoribbons, J. Phys. Chem. Lett. 3 (2012) 924–929. https://doi.org/10.1021/jz300164p.

[32] Y.-C. Chen, D.G. de Oteyza, Z. Pedramrazi, C. Chen, F.R. Fischer, M.F. Crommie, Tuning the Band Gap of Graphene Nanoribbons Synthesized from Molecular Precursors, ACS Nano. 7 (2013) 6123–6128. https://doi.org/10.1021/nn401948e.

[33] S. Linden, D. Zhong, A. Timmer, N. Aghdassi, J.H. Franke, H. Zhang, X. Feng, K. Müllen, H. Fuchs, L. Chi, H. Zacharias, Electronic Structure of Spatially Aligned Graphene Nanoribbons on Au(788), Phys. Rev. Lett. 108 (2012) 216801. https://doi.org/10.1103/PhysRevLett.108.216801.

[34] M. Di Giovannantonio, O. Deniz, J.I. Urgel, R. Widmer, T. Dienel, S. Stolz, C. Sánchez-Sánchez, M. Muntwiler, T. Dumslaff, R. Berger, A. Narita, X. Feng, K. Müllen, P. Ruffieux, R. Fasel, On-Surface Growth Dynamics of Graphene Nanoribbons: The Role of Halogen Functionalization, ACS Nano. 12 (2018) 74–81. https://doi.org/10.1021/acsnano.7b07077.

[35] P.P. Shinde, J. Liu, T. Dienel, O. Gröning, T. Dumslaff, M. Mühlinghaus, A. Narita, K. Müllen, C.A. Pignedoli, R. Fasel, P. Ruffieux, D. Passerone, Graphene nanoribbons with mixed cove-cape-zigzag edge structure, Carbon. 175 (2021) 50–59. https://doi.org/10.1016/j.carbon.2020.12.069.

[36] J. Overbeck, G. Borin Barin, C. Daniels, M.L. Perrin, L. Liang, O. Braun, R. Darawish, B. Burkhardt, T. Dumslaff, X.-Y. Wang, A. Narita, K. Müllen, V. Meunier, R. Fasel, M. Calame, P. Ruffieux, Optimized Substrates and Measurement Approaches for Raman Spectroscopy of Graphene Nanoribbons, Phys. Status Solidi B. 256 (2019) 1900343. https://doi.org/10.1002/pssb.201900343.

[37] G. Borin Barin, A. Fairbrother, L. Rotach, M. Bayle, M. Paillet, L. Liang, V. Meunier, R. Hauert, T. Dumslaff, A. Narita, others, Surface-Synthesized Graphene Nanoribbons for Room Temperature Switching Devices: Substrate Transfer and ex Situ Characterization, ACS Appl. Nano Mater. 2 (2019) 2184–2192.

[38] J. Overbeck, G.B. Barin, C. Daniels, M.L. Perrin, O. Braun, Q. Sun, R. Darawish, M. De Luca, X.-Y. Wang, T. Dumslaff, A. Narita, K. Müllen, P. Ruffieux, V. Meunier, R. Fasel, M. Calame, A Universal Length-Dependent Vibrational Mode in Graphene Nanoribbons, ACS Nano. 13 (2019) 13083–13091. https://doi.org/10.1021/acsnano.9b05817.

[39] I.A. Verzhbitskiy, M.D. Corato, A. Ruini, E. Molinari, A. Narita, Y. Hu, M.G. Schwab, M. Bruna, D. Yoon, S. Milana, X. Feng, K. Müllen, A.C. Ferrari, C. Casiraghi, D. Prezzi, Raman Fingerprints of Atomically Precise Graphene Nanoribbons, Nano Lett. 16 (2016) 3442–3447. https://doi.org/10.1021/acs.nanolett.5b04183.

[40] G. Borin Barin, Q. Sun, M. Di Giovannantonio, C. Du, X. Wang, J.P. Llinas, Z. Mutlu, Y. Lin, J. Wilhelm, J. Overbeck, C. Daniels, M. Lamparski, H. Sahabudeen, M.L. Perrin, J.I. Urgel, S. Mishra, A.

Kinikar, R. Widmer, S. Stolz, M. Bommert, C. Pignedoli, X. Feng, M. Calame, K. Müllen, A. Narita, V. Meunier, J. Bokor, R. Fasel, P. Ruffieux, Growth Optimization and Device Integration of Narrow-Bandgap Graphene Nanoribbons, Small. (2022) 2202301. https://doi.org/10.1002/smll.202202301.

[41] A. Jorio, R. Saito, G. Dresselhaus, M.S. Dresselhaus, Raman Spectroscopy in Graphene Related Systems, Wiley-VCH Verlag GmbH & Co. KGaA, 2011. http://onlinelibrary.wiley.com/doi/10.1002/9783527632695.fmatter/summary (accessed November 16, 2015).

[42] F. Léonard, Crosstalk between nanotube devices: contact and channel effects, Nanotechnology. 17 (2006) 2381. https://doi.org/10.1088/0957-4484/17/9/051.

[43] T. Classen, G. Fratesi, G. Costantini, S. Fabris, F.L. Stadler, C. Kim, S. de Gironcoli, S. Baroni, K. Kern, Templated Growth of Metal-Organic Coordination Chains at Surfaces, Angew. Chem. Int. Ed. 44 (2005) 6142–6145. https://doi.org/10.1002/anie.200502007.

[44] M.E. Cañas-Ventura, W. Xiao, D. Wasserfallen, K. Müllen, H. Brune, J.V. Barth, R. Fasel, Self-Assembly of Periodic Bicomponent Wires and Ribbons, Angew. Chem. Int. Ed. 46 (2007) 1814–1818. https://doi.org/10.1002/anie.200604083.

[45] F. Tuinstra, Raman Spectrum of Graphite, J. Chem. Phys. 53 (1970) 1126. https://doi.org/10.1063/1.1674108.

[46] M.S. Dresselhaus, A. Jorio, M. Hofmann, G. Dresselhaus, R. Saito, Perspectives on Carbon Nanotubes and Graphene Raman Spectroscopy, Nano Lett. 10 (2010) 751–758. https://doi.org/10.1021/nl904286r.

[47] A. Maghsoumi, L. Brambilla, C. Castiglioni, K. Müllen, M. Tommasini, Overtone and combination features of G and D peaks in resonance Raman spectroscopy of the C78H26 polycyclic aromatic hydrocarbon, J. Raman Spectrosc. 46 (2015) 757–764. https://doi.org/10.1002/jrs.4717.

[48] M. Tommasini, C. Castiglioni, G. Zerbi, Raman scattering of molecular graphenes, Phys. Chem. Chem. Phys. 11 (2009) 10185. https://doi.org/10.1039/b913660f.

[49] R. Gillen, M. Mohr, C. Thomsen, J. Maultzsch, Vibrational properties of graphene nanoribbons by first-principles calculations, Phys. Rev. B. 80 (2009). https://doi.org/10.1103/PhysRevB.80.155418.

[50] R. Gillen, M. Mohr, J. Maultzsch, Symmetry properties of vibrational modes in graphene nanoribbons, Phys. Rev. B. 81 (2010). https://doi.org/10.1103/PhysRevB.81.205426.

[51] D. Liu, C. Daniels, V. Meunier, A.G. Every, D. Tománek, In-plane breathing and shear modes in low-dimensional nanostructures, Carbon. 157 (2020) 364–370. https://doi.org/10.1016/j.carbon.2019.10.041.

[52] Y. Wang, Y. Zheng, X. Xu, E. Dubuisson, Q. Bao, J. Lu, K.P. Loh, Electrochemical Delamination of CVD-Grown Graphene Film: Toward the Recyclable Use of Copper Catalyst, ACS Nano. 5 (2011) 9927–9933. https://doi.org/10.1021/nn203700w.

[53] L. Gao, W. Ren, H. Xu, L. Jin, Z. Wang, T. Ma, L.-P. Ma, Z. Zhang, Q. Fu, L.-M. Peng, X. Bao, H.-M. Cheng, Repeated growth and bubbling transfer of graphene with millimetre-size single-crystal grains using platinum, Nat. Commun. 3 (2012) 699. https://doi.org/10.1038/ncomms1702.

[54] H.H. Gommans, J.W. Alldredge, H. Tashiro, J. Park, J. Magnuson, A.G. Rinzler, Fibers of aligned single-walled carbon nanotubes: Polarized Raman spectroscopy, J. Appl. Phys. 88 (2000) 2509–2514. https://doi.org/10.1063/1.1287128.

[55] C. Kramberger, T. Thurakitseree, S. Chiashi, E. Einarsson, S. Maruyama, On the polarization-dependent Raman spectra of aligned carbon nanotubes, Appl. Phys. A. 109 (2012) 509–513. https://doi.org/10.1007/s00339-012-7305-8.

[56] J. Hwang, H.H. Gommans, A. Ugawa, H. Tashiro, R. Haggenmueller, K.I. Winey, J.E. Fischer, D.B. Tanner, A.G. Rinzler, Polarized spectroscopy of aligned single-wall carbon nanotubes, Phys. Rev. B. 62 (2000) R13310–R13313. https://doi.org/10.1103/PhysRevB.62.R13310.

[57] M. Ichida, S. Mizuno, H. Kataura, Y. Achiba, A. Nakamura, Anisotropic optical properties of mechanically aligned single-walled carbon nanotubes in polymer, Appl. Phys. A. 78 (2004) 1117–1120. https://doi.org/10.1007/s00339-003-2462-4.

[58] A.K. Gupta, T.J. Russin, H.R. Gutiérrez, P.C. Eklund, Probing Graphene Edges *via* Raman Scattering, ACS Nano. 3 (2009) 45–52. https://doi.org/10.1021/nn8003636.

[59] Z. Li, R.J. Young, I.A. Kinloch, N.R. Wilson, A.J. Marsden, A.P.A. Raju, Quantitative determination of the spatial orientation of graphene by polarized Raman spectroscopy, Carbon. 88 (2015) 215–224. https://doi.org/10.1016/j.carbon.2015.02.072.

[60] P. Kannan, S.J. Eichhorn, R.J. Young, Deformation of isolated single-wall carbon nanotubes in electrospun polymer nanofibres, Nanotechnology. 18 (2007) 235707. https://doi.org/10.1088/0957-4484/18/23/235707.

# Supporting Information

**Quantifying Alignment and Quality of Graphene Nanoribbons: A Polarized Raman Spectroscopy Approach**

Rimah Darawish,[1, 2] Jan Overbeck,[3, 4] Klaus Müllen,[5] Michel Calame,[3, 4, 6] Pascal Ruffieux,[1] Roman Fasel,[1, 2] and Gabriela Borin Barin[1]

[1]*Empa, Swiss Federal Laboratories for Materials Science and Technology, nanotech@surfaces Laboratory, 8600 Dübendorf, Switzerland*

[2]*Department of Chemistry, Biochemistry and Pharmaceutical Sciences, University of Bern, 3012 Bern, Switzerland*

[3]*Empa, Swiss Federal Laboratories for Materials Science and Technology, Transport at Nanoscale Interfaces Laboratory, 8600 Dübendorf, Switzerland*

[4]*Swiss Nanoscience Institute, University of Basel, 4056 Basel, Switzerland*

[5]*Max Planck Institute for Polymer Research, 55128 Mainz, Germany*

[6]*Department of Physics, University of Basel, 4056 Basel, Switzerland*

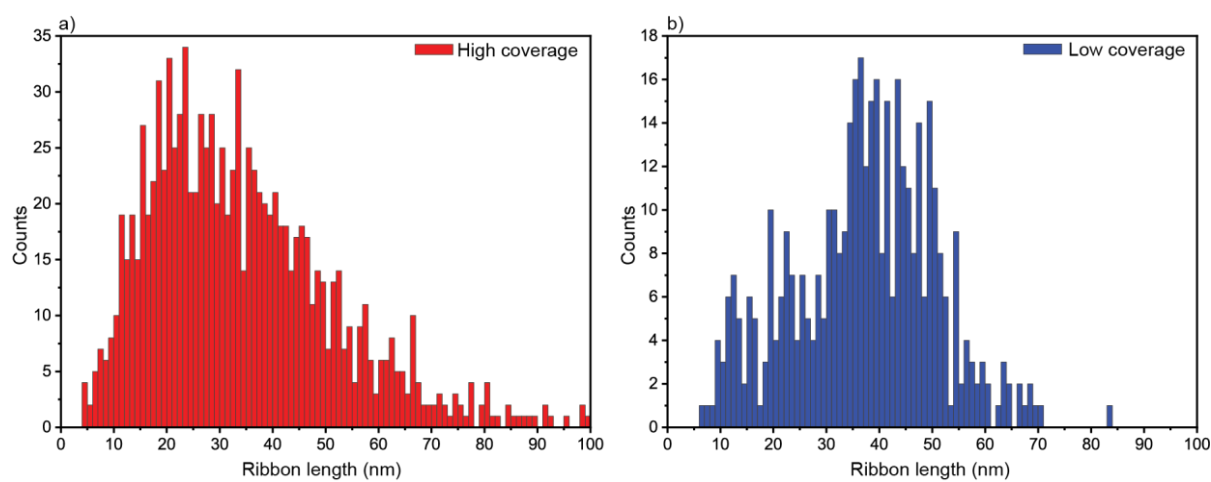

Figure S1: Length Distribution of 9-AGNRs for High- and Low-Coverage Samples on Au(788). (a) Histogram of GNR lengths for the high-coverage sample, with an average length of 34 nm. The histogram is generated by analyzing 5 representative STM images, encompassing a total of 1051 GNRs. (b) Histogram of GNR lengths for the low-coverage sample, with an average length of 37 nm. The histogram is generated by analyzing 15 representative STM images, totaling 416 GNRs. GNRs shorter than 4 nm and ill-defined structures are excluded from the analysis.

## The influence of quality of alignment ($\sigma$) and overall disorder on the surface (*OD*) on polarized Raman intensity and Raman polarization anisotropy (*P*)

To understand the effect of $\sigma$ (represented by the width of the GNR angle distribution) and *OD* on the Raman intensity, let's have a closer look at the meaning and impact of these parameters. Precisely speaking, $\sigma$ is the standard deviation of the Gaussian angle distribution centered around the mean azimuthal direction $\vartheta_0$ and is directly related to the $\text{FWHM} = 2\sigma \times \sqrt{2 \times \ln(2)} = 2.35\sigma$. We thus assume the GNRs to be aligned "on average" along an azimuthal direction $\vartheta_0$, with the distribution of GNR angles

# Supporting Information

described by a Gaussian distribution $G(\vartheta) = \frac{1}{\sigma\sqrt{2\pi}} e^{-\frac{(\vartheta-\vartheta_0)^2}{2\sigma^2}}$ as defined in Eq. (2). The *OD* parameter defines the ratio of the isotropic (polarization-independent) contribution (*B*) to the total Raman intensity (*A+B*), where *A* is the polarization-dependent Raman intensity resulting from the aligned GNRs. More precisely, if *A* and *B* are the fractional (*A+B*=1) surface areas giving rise to polarization-dependent and polarization-independent Raman intensities, respectively, then *OD* = *B/(A+B)* = *B*.

The degree of alignment is frequently characterized by the value of the polarization anisotropy $P = (I_{max} - I_{min})/(I_{max} + I_{min})$ determined from the Raman intensity $I_{max}$ measured along the preferred angle direction $\vartheta_0$ and $I_{min}$ measured orthogonal to it, but this implies that $\vartheta_0$ is known, which is not generally the case. Also, the determination of *P* in this way does not allow to distinguish between low polarization anisotropy due to a broad distribution of GNR angles (low degree of uniaxial alignment) or due to an important contribution of polarization-independent Raman intensity stemming from contributions other than preferentially aligned GNRs (such as polymer residues, contaminants, randomly oriented unreacted precursor molecules, etc.). Therefore, measuring the polarization dependence of the Raman intensity $I_{exp}(\vartheta)$ over a 180° range of azimuthal angles $\vartheta$ provides much more detailed information, with which both contributions discussed above can be disentangled.

Practically, the experimental data of $I_{exp}(\vartheta)$ is fitted with Eq. (9),

$$I_{exp}(\vartheta) = A \cdot \int_{0°}^{360°} \cos^4(\varphi) \cdot \frac{1}{\sigma\sqrt{2\pi}} e^{-\frac{(\vartheta-\varphi-\vartheta_0)^2}{2\sigma^2}} d\varphi + B \cdot \frac{3}{8}$$

and the resulting fit parameters $\vartheta_0$, $\sigma$ and *B* (*A*=1-*B*) determine the direction of preferential GNR alignment ($\vartheta_0$), the degree of GNR alignment ($\sigma$), and the overall degree of disorder *OD* = *B*.

To understand the impact of $\sigma$ and *OD* on the polarized Raman intensity $I_{pol}(\vartheta)$, we plot in Fig. S2 the dependence of $I_{pol}(\vartheta)$ on both parameters individually. It can be seen that an increasing contribution of polarization-independent Raman intensity (as specified by *OD*) leads to a vertical offset of the $I_{pol}(\vartheta)$ curves, i.e. the minimum Raman intensity $I_{min}$ no longer drops to zero even for polarization directions perpendicular to the GNR axis. Increasing the width of the Gaussian distribution of GNR angles, on the other hand, leads to a broadening of $I_{pol}(\vartheta)$ and a quick decrease of the maximum Raman intensity $I_{max}$ measured along the preferential direction of alignment, whereas the minimum intensity $I_{min}$ measured in the perpendicular direction is significantly less affected.

The corresponding *P* values as a function of both $\sigma$ and *OD* are depicted in Fig S3. Increasing *OD* (*OD* > 0) immediately decreases *P* strongly, in an almost linear fashion. *P* is, however, much less sensitive to the width of the Gaussian distribution (as specified by $\sigma$), at least for values smaller than 10-15° where

# Supporting Information

*P* depends only weakly on *σ*. Larger *σ* (15-60°) then strongly decreases *P*. In other words, *P* is a reasonably good indicator of the combined impact of *σ* and *OD*, but rather insensitive to width differences of narrow (*σ* < 15°) angle distributions.

It is thus clear that measuring the full polarization dependence of the Raman intensity $I_{exp}(\vartheta)$ over a 180° range of azimuthal angles $\vartheta$ provides much more complete information than simply determining the polarization anisotropy $P = (I_{Max} - I_{min})/(I_{max} + I_{min})$ determined from the Raman intensities $I_{Max}$ measured along the (generally unknown) preferred angles direction $\vartheta_0$ and $I_{min}$ measured orthogonal to it. In particular, fitting $I_{exp}(\vartheta)$ with Eq. (9) allows for characterizing both the angular distribution of GNRs and the disordered surface area contributions.

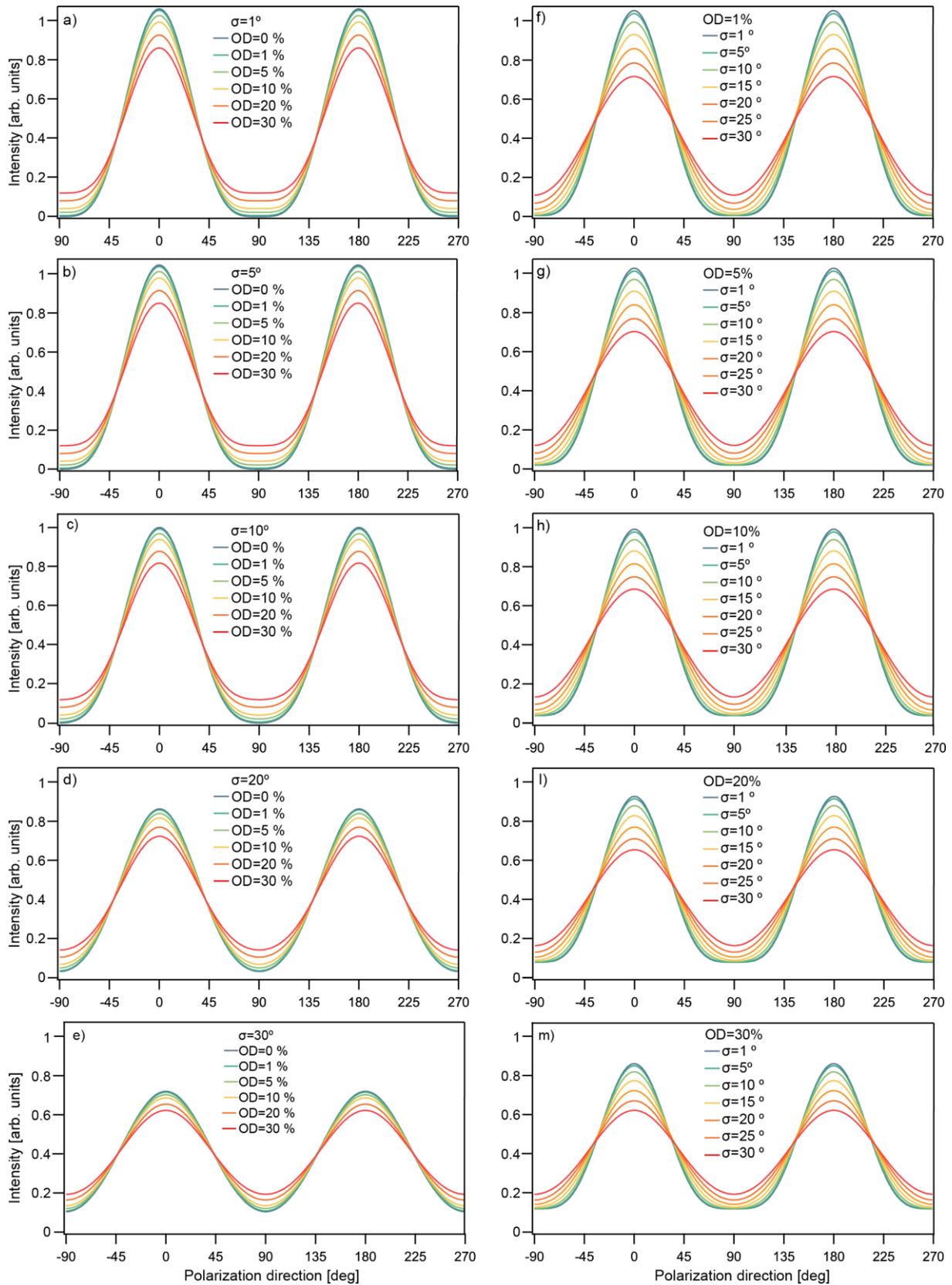

Figure S2: Polarized Raman intensity behavior with varying *σ* and *OD*.

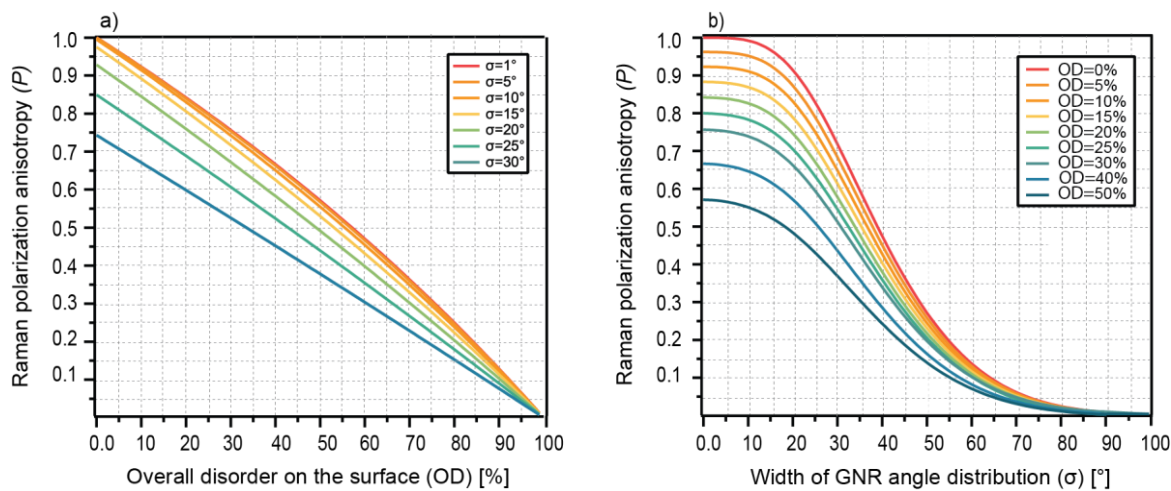

Figure S3: Impact of *OD* and *σ* on *P*. (a) Relationship between *P* and *OD* for different values of *σ*. (b) Relationship between *P* and *σ* for different values of *OD*.

**Polarization dependence of 9-AGNRs Raman active modes with VV configuration**

To explore the polarization dependence of the other Raman active modes of 9-AGNRs, we have also acquired polarized Raman intensity data as a function of polarization direction for the RBLM, CH, and D modes, both from the same high- and low-coverage 9-AGNR samples in the main text on the Au(788) growth substrate and after transfer to a ROS. The resulting data sets are shown in Figs S4 and S5 for the high- and low-coverage samples, respectively. We observed that CH, D, and RBLM modes exhibit roughly the same polarization dependence (VV configuration) as the G mode, with a maximum intensity when the incident polarization is parallel to the ribbon axis (0°, 180°) and minimum when perpendicular.

As discussed and shown in the main text for the G mode data, the extended polarization fitting model has also been applied to the RBLM, CH, and D peak Raman data. Tables S1 and S2 summarize the resulting *σ*, *OD*, and *P*-values for the high- and low-coverage samples on Au(788) and after substrate transfer onto the ROS, respectively. The *σ*, *OD*, and *P*-values of RBLM, CH, and D in Tables S1 and S2 exhibit comparable values to the G mode (presented in the main manuscript Table 1) for both high- and low-coverage samples on Au(788) and ROS, with a larger error range for low-coverage samples, which is due to the low Raman intensities for these samples.

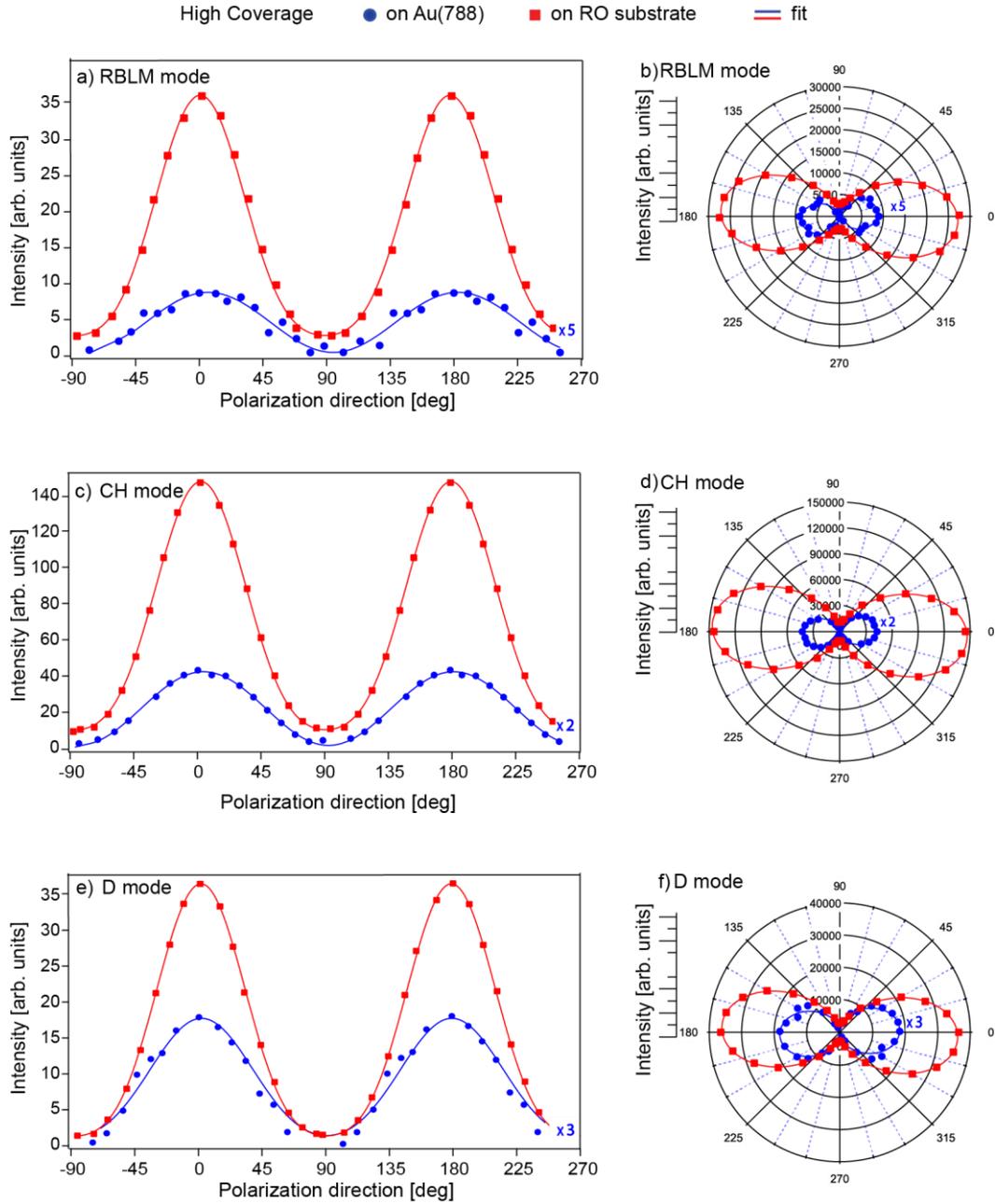

Figure S4: Polarized Raman intensity of RBLM, CH, and D modes for high-coverage sample. (a), (c), and (e) Measured RBLM, CH, and D mode intensities as a function of polarization angle $\vartheta$ on Au(788) (blue circles) and after substrate transfer onto ROS (red squares). Data fits using Eq. (9) are represented by blue and red solid lines. (b), (d), and (f) Representative polar diagrams illustrating the RBLM, CH, and D mode intensities on Au(788) (blue circles) and ROS (red squares), corresponding to the same data as in (a), (c), and (e), respectively. Blue and red solid lines represent the fits to the measured data using Eq. (9).

# Supporting Information

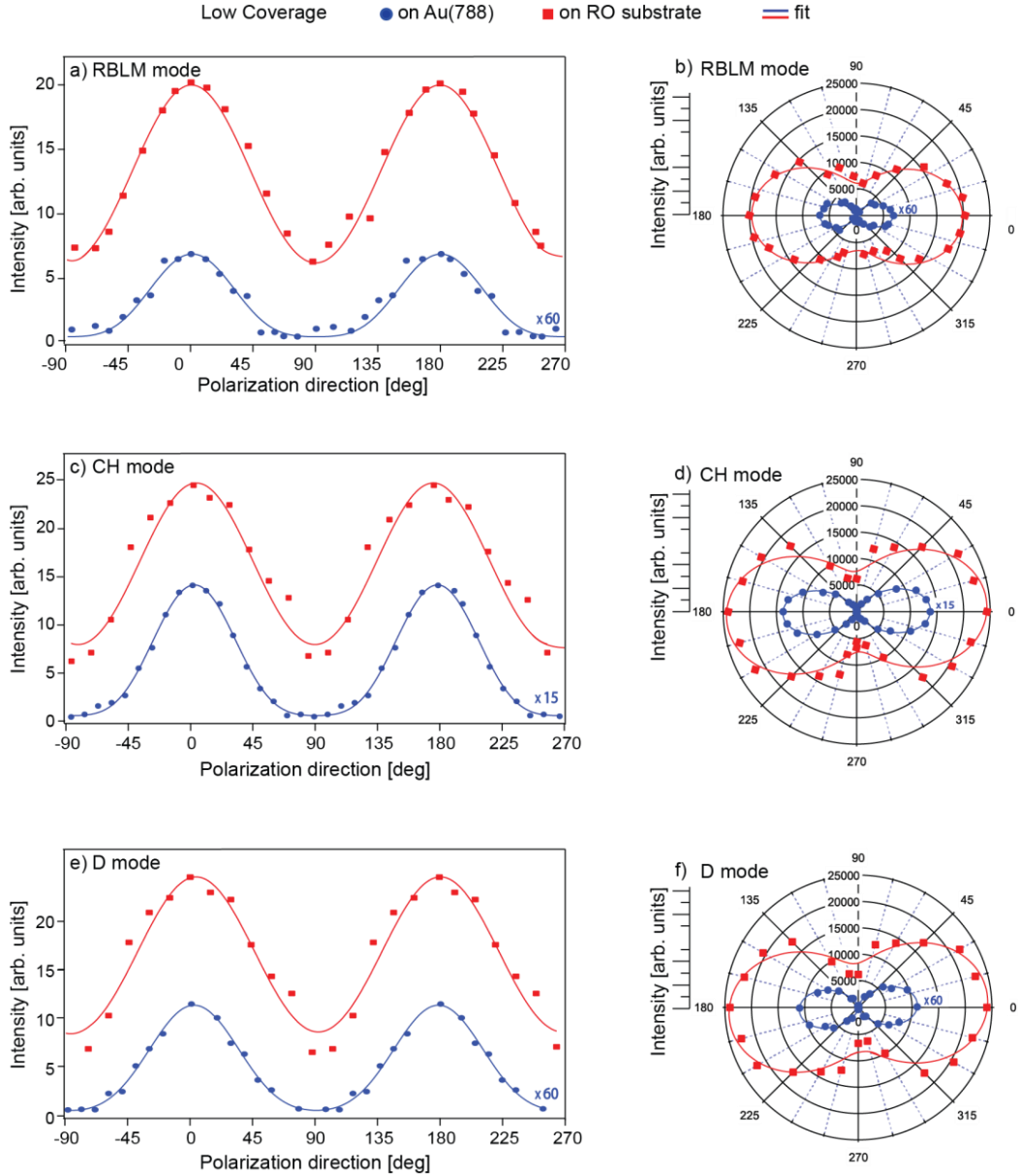

Figure S5: Polarized Raman intensity of RBLM, CH, and D modes for low-coverage sample. (a), (c), and (e) Measured RBLM, CH, and D mode intensities as a function of polarization angle $\vartheta$ on Au(788) (blue circles) and after substrate transfer onto ROS (red squares). Data fits using Eq. (9) are represented by blue and red solid lines. (b), (d), and (f) Representative polar diagrams illustrating the RBLM, CH, and D mode intensities on Au(788) (blue circles) and ROS (red squares), corresponding to the same data as in (a), (c), and (e), respectively. Blue and red solid lines represent the fits to the measured data using Eq. (9).

# Supporting Information

| High-coverage sample | RBLM peak | | CH peak | | D peak | |
|---|---|---|---|---|---|---|
| | Au(788) | ROS | Au(788) | ROS | Au(788) | ROS |
| Quality of alignment ($\sigma$) [°] | 3± 1 | 13± 1 | 3± 1 | 15± 2 | 2± 1 | 14± 1 |
| Overall disorder on the surface ($OD$) [%] | 8± 1 | 14± 1 | 7± 1 | 13± 1 | 8± 2 | 14± 1 |
| Raman polarization anisotropy ($P$) | 0.86 | 0.81 | 0.86 | 0.84 | 0.86 | 0.85 |

Table S1: High-coverage 9-AGNR sample $\sigma$, $OD$, and $P$-values resulting from the fitting of the RBLM, CH, and D mode data with the 9-AGNRs on Au(788) and after transfer to the ROS.

| Low-coverage sample | RBLM peak | | CH peak | | D peak | |
|---|---|---|---|---|---|---|
| | Au(788) | ROS | Au(788) | ROS | Au(788) | ROS |
| Quality of alignment ($\sigma$) [°] | 1.0± 0.3 | 22± 4 | 1.1± 0.2 | 26± 4 | 1.5± 0.4 | 26± 1 |
| Overall disorder on the surface ($OD$) [%] | 8± 4 | 48± 6 | 8± 2 | 39± 3 | 6± 4 | 43± 5 |
| Raman polarization anisotropy ($P$) | 0.91 | 0.54 | 0.90 | 0.50 | 0.90 | 0.55 |

Table S2: Low-coverage 9-AGNR sample $\sigma$, $OD$, and $P$-values resulting from the fitting of the RBLM, CH, and D mode data with the 9-AGNRs on Au(788) and after transfer to the ROS.

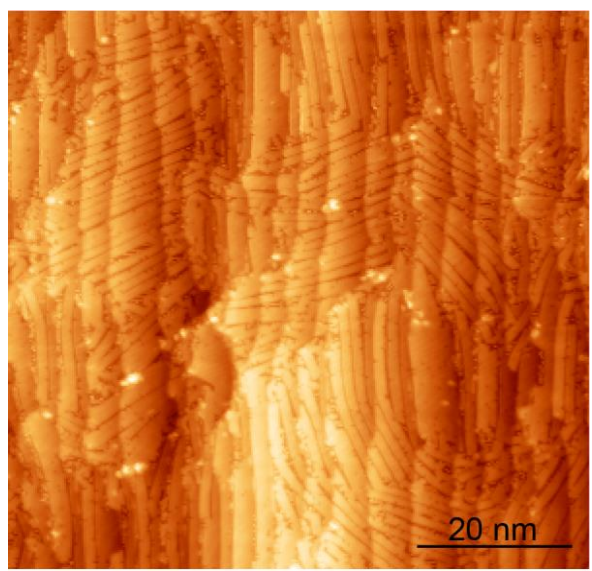

Figure S6: Constant current STM image of the high-coverage sample of 9-AGNRs on Au(788) taken at a location showing GNRs growing across the terraces ($V_b$ = -1.5 V, $I_t$ = 0.3nA, scale bar of 20 nm).